\documentclass{article}
\usepackage[utf8]{inputenc}
\usepackage[T1]{fontenc}

\usepackage{amsmath,amssymb,amsfonts}
\usepackage{graphicx}

\newtheorem{definition}{Definition}[section]
\newtheorem{theorem}{Theorem}[section]

\title{Two-step method for assessing dissimilarity of random sets}

\author{\footnote{$^1$ University of Split, Croatia; $^2$ Czech Technical University in Prague, Czech Republic ($^{\ddag}$ email: heliskat@fel.cvut.cz); $^3$ Charles University, Czech Republic; $^4$ University of Chemistry and Technology, Czech Republic} Vesna Gotovac \DJ oga\v{s}$^1$, Kate\v{r}ina Helisov\'a$^{2,\ddag}$, Bogdan Radovi\'c$^2$, Jakub Stan\v{e}k$^3$,\\ Mark\'eta Zikmundov\'a$^4$, Kate\v{r}ina Brejchov\'a$^2$}

\begin{document}

\maketitle

\begin{abstract}
The paper concerns a new statistical method for assessing dissimilarity of two random sets based on one realisation of each of them.
The method focuses on shapes of the components of the random sets, namely on the curvature of their boundaries together with the ratios of their perimeters and areas.
Theoretical background is introduced and then, the method is described, justified by a simulation study and applied to real data of two different types of tissue - mammary cancer and mastopathy.
\end{abstract}

\section{Introduction}

In the last years, modelling and statistical analyses of random sets have become very popular. 
It has been studied both from theoretical point of view 
\cite{matheron:1975}, \cite{molchanov:2005}, \cite{serra:1982} as well as from the practical side, because they have many applications in biology \cite{moeller:2010}, medicine \cite{hermann:2015}, material sciences \cite{neumann:2016} and other branches.
They can describe and explain many events, for example behaviour of cells in organisms, 
particles in materials, 
presence of different plants etc. 
Therefore mathematical methods dealing with random sets must be developed and improved.

Usually, when we are given a realisation of a random set, we try to find, based on the realisation, a model in order to make further statistical analyses. 
However, there are situations when the knowledge about the concrete model is not necessary, because the aim is to decide whether two realisations are similar in some sense, i.e. whether they may come from the same process, e.g. we need only to distinguish between two types of cells in tissue from microscopic pictures, recognise different tendency of growth of some plants, detect defects in materials etc.

In the presented paper, we focus on planar random sets, nevertheless, the method described here can be easily extended to more dimensional spaces.
Although there exist classical tools for comparing random sets like covariance function or 
contact distribution function~\cite{chiu:2013}, 
functions on morphological operations, namely dilation, erosion, opening and closing~\cite{serra:1982}, 
etc., there are situations when these characteristics are not sufficient to distinguish between two realisations. E.g. we can obtain the same estimate of contact distribution function for different shapes or, on the other hand, two realisations consisting of components of the same shapes but different mutual distances have very different estimate of the covariance function etc., so they cannot be used when the main objects of interest are the shapes of the components, as in this paper.
Another disadvantage of the mentioned characteristics is that for one realisation, we obtain only one function.
Then, it is difficult to formulate the task of comparing two random sets when we have only one realisation of each of them.
In this case, 
it would be helpful to have data consisting of more functions for each realisation.

New methods taking this into account have been developed in the last five years, see \cite{debayle:2021}, \cite{gotovac:2019}, \cite{gotovac:2021} and \cite{gotovac:2016}. 
In~\cite{debayle:2021}, morphological skeletons~\cite{serra:1982} of compared realisations are constructed and then, similarity of two realisations is defined through a function describing mass growth around selected points of the skeletons.
This method shows the highest power in simulation study, compared to the methods in the remaining three papers.
However, it is closely tied to the placement of components in realisations, which is not always desirable.
Further approach can be found in \cite{gotovac:2019}, where the author also focuses on similarity of shapes and positions of components in realisations, but the positions can be omitted in special cases. 
The considered components are either the connected components or some more specific set components that are obtained by further decomposing of the observed set (e.g. cells in a tissue which are connected but they can be determined in their binary image). 
The positions of the components are described by the so-called neighbourhood tessellation of the observation window constructed from the components and by using the Hausdorff metric~\cite{chiu:2013}. 
The samples of pairs of the components and neighbouring tessellation cells are compared using the test based on the $\mathcal N$-distance~\cite{klebanov:2006}, where a suitable kernel involving the symmetric difference of the sets is constructed.
For omitting dependence on locations, the components can be considered without their neighbourhoods.
A disadvantage of the method is that the test of similarity in this sense is weak, especially when the neighbourhoods are omitted, so in order to achieve better accuracy, the positions of the components must be taken into account again.
An approach independent of the positions of the components is introduced in~\cite{gotovac:2016} and improved in~\cite{gotovac:2021}. 
The authors of the papers distinguish between two realisations via a heuristic approach based on approximation of the components by unions of convex compact sets using Voronoi tessellation~\cite{chiu:2013} with respect to a special hard-core point process on the realisations
and consequent comparison of the support functions of the convex compact cells of the tessellation using the envelope test from \cite{myllymaki:2017} and the test based on $\mathcal N$-distance~\cite{klebanov:2006} 
(sometimes also called the kernel test, see \cite{gretton:2012}), respectively.
The authors of \cite{gotovac:2021} and \cite{gotovac:2016} declare that the method focuses on the structure of the components like clustering or repulsion tendencies, generating rounded or angular objects, long and thin or short and thick formations, etc., because it recognises how much mass is concentrated on the boundaries of the components and what is the approximate shape of the boundary.
The results of simulation study in \cite{gotovac:2021} and \cite{gotovac:2016} are very satisfactory, however, there are some weak parts in the heuristic procedure coming from the fact that the approximation of the components is not unique, but random, and there are quite large differences between the shapes of the original realisations and their approximations. 
Moreover, some free parameters must be chosen, which can affect the results. 

In the present paper, we focus on testing similarity of two realisations of random sets, while the similarity is given by similar shapes of their components, analogously to~\cite{gotovac:2021} and \cite{gotovac:2016}.
However, instead of the heuristic approximation of realisations, we use description by uniquely defined characteristics of components in the realisations, namely the ratio of the perimeter and the area of each component, and the curvature~\cite{bullard:1995} of the boundary of each component. 

The paper is organised as follows. 
Section \ref{sec:TB} summarises definitions and the already existing theoretical results concerning curvatures of planar curves and statistical testing via the $\mathcal{N}$-distance theory. 
In Section~\ref{sec:method}, we define similarity of two random sets and describe the procedure of assessing similarity from two realisations. 
This is the main result of the paper. 
Section~\ref{sec:simul} justifies the procedure from Section~\ref{sec:method} by a simulation study and compares the results to the results obtained by the previous methods.

\section{Theoretical background}
\label{sec:TB}

\subsection{Curvature of a planar curve}
\label{sec:curv}

Consider a smooth 2D curve $c$ parameterised by a parameter $\varphi \in [0,\phi] \subset \mathbb R$, i.e. $c(\varphi)=(x(\varphi),y(\varphi))$.
Then the curvature $\kappa$ of $c$ is defined as 
$$\kappa(c(\varphi)) = \frac{x'(\varphi)y''(\varphi)-x''(\varphi)y'(\varphi)}{(x'^2(\varphi)+y'^2(\varphi))^{3/2}},$$
see e.g.~\cite{bullard:1995}.
It means that $\kappa(c(\varphi)) = \pm 1/R(\varphi)$ , where $R(\varphi)$ is the radius of the osculating circle touching the curve in the point $[x(\varphi),y(\varphi)]$ and the choice between “$+$” and “$-$” is determined by the local convexity convention.

Let us assume that the curve $c$ is continuous, closed (i.e. $c(0)=c(\phi)$) and it does not intersect itself (i.e. $c(\varphi_1)=c(\varphi_2) \Rightarrow \varphi_1=\varphi_2$). 
Consider a (connected) planar set $X$ whose boundary is given by the curve $c$. 
It can be shown~\cite{bullard:1995} that for the curvature $\kappa(z)$ evaluated in a given point $z \in c$ and for a disc $b(z,r)$ with the center in $z$ and a radius $r$ small enough, it holds that
\begin{equation}
\label{eq:curv}   
\kappa(z) \approx \frac{3A^*_{b(z,r)}}{r^3}-\frac{3\pi}{2r} = \frac{3\pi}{r}\left(\frac{A^*_{b(z,r)}}{A_{b(z,r)}}-\frac{1}{2}\right),
\end{equation}
where $A_{b(z,r)}$ is the area of the disc $b(z,r)$ and $A^*_{b(z,r)}$ is the area of $b(z,r) \cap X$.

This is used below in Section~\ref{sec:method} when estimating the curvature of boundary of binary image of the set $X$.
The center of discretised disc $b$ with the radius $r$ pixels is placed to the boundary pixel in which we want to evaluate the curvature, and approximate the ratio $A^*_{b(z,r)}/A_{b(z,r)}$ by the number of pixels of the disc $b$ inside the set $X$ divided by the number of all pixels forming the disc $b$.

\subsection{Testing equality in distribution based on $\mathcal{N}$-distance of probability measures}
\label{sec:Ndist}

In this paper, testing equality in distribution of random variables as well as random functions comes from the theory of $\mathcal{N}$-distances (see~\cite{klebanov:2006}) that is briefly recalled in the following paragraphs.

Let $\mathcal X$ be a nonempty set.
Consider a negative definite kernel $\mathcal L: \mathcal X \times \mathcal X \rightarrow \mathbb C$ (see \cite[pp.~16--19]{klebanov:2006}).

\begin{definition}  
The negative definite kernel 
$\mathcal L$ is called \emph{strongly negative definite kernel} if 
for an arbitrary probability measure $\mu$ and an arbitrary $f:\mathcal X \to \mathbb R$ such that $\int_{\mathcal X} f(x) d\mu( x)=0$ holds and $\int_{\mathcal X} \int_{\mathcal X} \mathcal L(x,y) f(x) f(y) d \mu (x) d \mu (y)$ exists  and is finite, the relation 
$$
\int_{\mathcal X} \int_{\mathcal X}   \mathcal L(x,y) 
f(x) f(y) d\mu( x) d\mu( y) = 0
$$
implies that $f(x)=0$ $\mu$-a.e.
\end{definition}


For a map $\mathcal L: \mathcal X \times \mathcal X \rightarrow \mathbb C$, denote $B_{\mathcal L}$ the set of all measures $\mu$ such that $\int_{\mathcal X} \int_{\mathcal X} \mathcal L(x,y) d\mu (x) d\mu (y)$ exists.

\begin{theorem}[Klebanov, 2006]
Let $\mathcal L(x,y) = \mathcal L(y,x)$. Then
\begin{align}
\label{eq:Ndist}
\mathcal N(\mu,\nu) = & 2\int_{\mathcal X} \int_{\mathcal X} \mathcal L(x,y) d\mu (x) d\nu (y) - 
\int_{\mathcal X} \int_{\mathcal X} \mathcal L(x,y) d\mu (x) d\mu (y) 
-\int_{\mathcal X} \int_{\mathcal X} \mathcal L(x,y) d\nu (x) d\nu (y)
\geq 0
\end{align}
holds for all measures $\mu, \nu \in \mathcal B_{\mathcal L}$ with 
equality in the case $\mu=\nu$ only, if and only if $\mathcal L$ is a strongly negative definite kernel.
\end{theorem}

In the sequel, the term $\mathcal N(\mu,\nu)$ from \eqref{eq:Ndist} is called the $\mathcal N$-distance of the measures $\mu$ and $\nu$.

Some examples of strongly negative definite kernels $\mathcal L$, which can be used for testing equality in distribution of real random variables, can be found in \cite[pp.~21--22]{klebanov:2006}. 
Here, we use the Euclidean distance as negative definite kernel when comparing only random values and a special kernel developed for comparing random functions described below in this section when comparing random functions.
In both cases, we apply a permutation version of test.
The approach is as follows.
Assume we have an observation $x_1,\ldots,x_{m_1}$ from an distribution $\mu$ and  $y_1,\ldots,y_{m_2}$ from an distribution $\nu$.
The $\mathcal N$-distance of the measures $\mu$ and $\nu$ is estimated as
\begin{align}\hat{\mathcal{N}_1}= \ &\frac{2}{m_1 m_2}\sum\limits_{i=1}^{m_1}\sum\limits_{j=1}^{m_2}\mathcal{L}(x_i,y_j)
-\frac{1}{m_1^2}\sum\limits_{i=1}^{m_1}\sum\limits_{j=1}^{m_1}\mathcal{L}(x_i,x_j)-\frac{1}{m_2^2}\sum\limits_{i=1}^{m_2}\sum\limits_{j=1}^{m_2}\mathcal{L}(y_i,y_j).
\label{eq:Nest1}
\end{align}
This value plays the role of the test characteristic.
Then, we use a Monte Carlo permutation test, i.e. we make $s$ permutations of all observed values $x_1, \ldots, x_{m_1}, y_1, \ldots y_{m_2}$, split each permutation into two groups of the lengths $m_1$ and $m_2$, and, analogously to \eqref{eq:Nest1}, we calculate $\hat{\mathcal{N}}_i$ for the $i$-th permutation, $i = 2,\ldots,s+1$.
Then the $p$-value of the test is
\begin{equation}
p=\frac{\sharp\{i \in \{2,\ldots,s+1\}:  \hat{\mathcal{N}_i} \geq \hat{\mathcal{N}_1}\} +1}{s+1}.
\label{eq:pval2}
\end{equation}

For testing the equality in distribution of random functions is Section~\ref{sec:method}, we use a kernel from \cite{gotovac:2021} constructed especially for random functions.
The test works as follows.
When we evaluate the testing functions $t^{(1)}$ and $t^{(2)}$ in discrete arguments $u_1,\ldots,u_n,$ $n \in \mathbb{N}$,
the kernel is
\begin{equation}
\mathcal{L}(t^{(1)},t^{(2)})=\sum\limits_{m=1}^{D}\sum\limits_{\left\{k_1,\ldots,k_m\right\} 
\subseteq \left\{1,\ldots,n\right\} }\left(\sum\limits_{l=1}^m \left(t^{(1)}(u_{k_l})-t^{(2)}(u_{k_l}))\right)^2 \right)^{1/2},
\label{eq:l_D}
\end{equation}
where $D$ is a chosen constant specifying the depth of dependence, i.e. it is suitable for testing the equality of finite-dimensional distributions of random functions $t^{(1)}$ and $t^{(2)}$ for the dimensions less than or equal to $D$ (for more details see \cite{gotovac:2021}). 
The choice of the kernel \eqref{eq:l_D} is based on the simulation study from \cite{gotovac:2021} which shows that it gives the largest test power from all studied kernels.
The estimate of the $\mathcal N$-distance of the functions $t^{(1)}$ and $t^{(2)}$ based on the random samples $t^{(1)}_i$, $i=1,\ldots,m_1$, and $t^{(2)}_j$, $j=1,\ldots,m_2$, respectively, is evaluated analogously to the previous, i.e. as
\begin{align}\hat{\mathcal{N}_1}= \ &\frac{2}{m_1 m_2}\sum\limits_{i=1}^{m_1}\sum\limits_{j=1}^{m_2}\mathcal{L}(t^{(1)}_i,t^{(2)}_j)
-\frac{1}{m_1^2}\sum\limits_{i=1}^{m_1}\sum\limits_{j=1}^{m_1}\mathcal{L}(t^{(1)}_i,t^{(1)}_j)
-\frac{1}{m_2^2}\sum\limits_{i=1}^{m_2}\sum\limits_{j=1}^{m_2}\mathcal{L}(t^{(2)}_i,t^{(2)}_j).
\label{eq:Nest2}
\end{align}
Then, we again use a Monte Carlo permutation test making $s$ permutations of all functions\break $t^{(1)}_1 (u), \ldots, t^{(1)}_{m_1} (u), t^{(2)}_1 (u), \ldots t^{(2)}_{m_2} (u)$ in order to obtain $\hat{\mathcal{N}}_i$, $i = 2,\ldots,s+1$, and evaluate the $p$-value as described above.

\section{Methodology}
\label{sec:method}

\subsection{Testing characteristics}

In this paper, we define the similarity of random sets through their components, namely through the curvature of their boundaries and the ratios of their perimeters and areas.

Consider a connected random set $\mathbf X$, i.e. the random set whose realisations are connected.
Denote $B_{\mathbf X}$ the boundary of $\mathbf X$ and $\kappa_{\mathbf X}(z)$ the (random) curvature in the point $z \in B_{\mathbf X}$.
From \eqref{eq:curv}, we can see that for a disc $b(z,r)$ with suitable chosen radius r, 
$$\kappa_{\mathbf X}(z) \propto \frac{A^*_{b(z,r),\mathbf X}}{A_{b(z,r)}},$$ 
where $A_{b(z,r)}$ is the area of the disc $b(z,r)$ and $A^*_{b(z,r),\mathbf X}$ is the area of $b(z,r) \cap \mathbf X$.
Therefore, we focus only on the ratio of these two areas. Denote
$$O_{\mathbf X,b(z,r)}=\frac{A^*_{b(z,r),\mathbf X}}{A_{b(z,r)}}$$
and define the function $$\tilde\kappa_{\mathbf X, r}(u) = |B_{\mathbf X}|^{-1}\int_{B_{\mathbf X}} \mathbf 1\{O_{\mathbf X,b(z,r)} \leq u \}dz, \quad u\in \langle 0,1\rangle,$$
which is basically an analogy of the distribution function of the curvature at points on the boundary, but it is evaluated for all boundary points, so it describes the distribution for strongly dependent values.
The object of our interest is the function, analogous to density function, describing the distribution of the curvature along the boundary, i.e.
\begin{equation}
\label{eq:ourcurv}
t_{\mathbf X, r}(u) = \tilde\kappa_{\mathbf X, r}'(u).
\end{equation}

Finally, denote $R_{\mathbf X}$ the random variable describing the ratio of the perimeter and the area of the random set $\mathbf X$.

\begin{definition}
\label{def:similar}
Two connected random sets $\mathbf X$ and $\mathbf Y$ are considered to be similar if the distributions of $\lim_{r\rightarrow 0} t_{\mathbf X, r}$ and $\lim_{r\rightarrow 0} t_{\mathbf Y, r}$ as well as the distributions of $R_{\mathbf X}$ and $R_{\mathbf Y}$ are equal.
\end{definition}

In practice, we usually observe realisations $X$ and $Y$ of the random sets $\mathbf X$ and $\mathbf Y$, respectively, in the form of binary images, so we need to adjust the task of assessing dissimilarity of the realisations consisting of black and white pixels.
The pixels play the role of units in the sequel.
The ratio of the perimeter and the area is then simply given by the number of boundary pixels divided by the number of all pixels of the component.
For evaluating of the function describing the curvature, fix a radius $r\in \mathbb N$, denote $P$ the set of all pixels of the binary image $X$, $z_1,\ldots,z_{n}$ all boundary pixels, and for each boundary pixel $z_i$, define
$$K(z_i)=\frac{\sharp \{p \in P: p \in b(z_i,r) \cap X\}}{\sharp \{p \in P: p \in b(z_i,r)\}}.$$
Then the approximation of the function~\eqref{eq:ourcurv} is
\begin{equation}
\label{eq:ourcurvest}
t(u)=\frac{\sharp\{i \in \{1,\ldots,n\}: K(z_i)\in[u-1/l,u)\}}{n}, \quad u=\frac{1}{l},\frac{2}{l},\ldots, 1,
\end{equation}
which plays the role of testing function.

\subsection{Testing similarity of connected random sets}
\label{sec:testconnect}

Consider two samples, namely $X_1,\ldots,X_{m_1}$ and $Y_1,\ldots,Y_{m_2}$, of realisations of connected random sets $\mathbf X$ and $\mathbf Y$, respectively.
We want to test the null hypothesis that $\mathbf X$ and $\mathbf Y$ are similar.
First we evaluate the ratios $R_{X_1},\ldots,R_{X_{m_1}}$, $R_{Y_1},\ldots,R_{Y_{m_2}}$ of the perimeters and areas of the corresponding realisations.
Based on these values, we estimate the $\mathcal N$-distance of the ratios of $\mathbf X$ and $\mathbf Y$ by the formula~\eqref{eq:Nest1}, where we set $x_i = R_{X_i}$, $i=1,\ldots,m_1$ and $y_j = R_{X_j}$, $j=1,\ldots,m_2$.
Let us denote it $\hat{\mathcal N^R_1}$.
Then we evaluate the testing functions~\eqref{eq:ourcurvest} describing the boundary curvatures $t_{X_1}(u),\ldots,t_{X_{m_1}}(u)$, $t_{Y_1}(u),\ldots,t_{Y_{m_2}}(u)$, calculate the $N$-distance of the functions corresponding to $\mathbf X$ and $\mathbf Y$, respectively, by the formulas~\eqref{eq:l_D} and ~\eqref{eq:Nest2}, and denote this $\mathcal N$-distance as $\hat{\mathcal N^t_1}$.
The couple $(\hat{\mathcal N^R_1},\hat{\mathcal N^t_1})$ is the test statistic.
Here, we use the Monte Carlo permutation test as described in Section~\ref{sec:Ndist}, i.e. we make $s$ permutations of all realisations $X_1,\ldots,X_{m_1}$ and $Y_1,\ldots,Y_{m_2}$, and split them into two groups of the sizes $m_1$ and $m_2$, respectively, in order to obtain $(\hat{\mathcal{N}}^R_i, \hat{\mathcal{N}}^t_i)$, $i = 2,\ldots,s+1$, and evaluate the $p$-value as 
\begin{equation}
p=\frac{\sharp\{i \in \{2,\ldots,s+1\}:  \hat{\mathcal{N}_i^R} \geq \hat{\mathcal{N}_1^R} \wedge \hat{\mathcal{N}_i^t} \geq \hat{\mathcal{N}_1^t}\} +1}{s+1}.
\label{eq:pval3}
\end{equation}

\subsection{Similarity of random sets consisting of more components}

Usually in practice, we have the data in the form of realisations consisting of more than one component.
If we can suppose that the components are independent and come from the same distribution, then we can define similarity of two random sets it the way that they are considered to be similar, if their components are similar in the meaning of Definition~\ref{def:similar}.
It is used in simulation study below in Section~\ref{sec:simul}.
Nevertheless, the independence of the components can be supposed in very specific cases, e.g. in 
germ-grain models~\cite{chiu:2013} in which the intensity of germs is low with respect to the volume of grains.
However, the components are usually dependent.
In order to avoid this complication, we make a suitable random sample of the components in each realisation, and use these two samples as the input samples $X_1,\ldots,X_{m_1}$ and $Y_1,\ldots,Y_{m_2}$ from Section~\ref{sec:testconnect}.

\section{Simulation study}
\label{sec:simul}

In the simulation study, we first focus on four models which illustrates the usage of the procedure. 
Then, we study models compared before in~\cite{debayle:2021}, \cite{gotovac:2019}, \cite{gotovac:2021} and \cite{gotovac:2016}.
The approach in the simulation study is analogous to that one presented in~\cite{debayle:2021}, \cite{gotovac:2019}, \cite{gotovac:2021} and \cite{gotovac:2016}.
For all considered models, we simulate 200 realisations and compare 100 vs 100 realisations of the same models as well as 100 vs 100 realisations of different models. 
Since the outputs of the tests are the $p$-values, we obtain 100 $p$-values for each couple. 
Some of their histograms are shown and commented below.
Note that $p$-value close to zero means that the equality of distributions of the corresponding testing functions is rejected. 
Thus, the $p$-value should be concentrated close to zero when comparing realisations of different models, while it should be uniformly distributed in the interval $[0,1]$ when comparing realisations of the same models.

In this simulation study, we moreover have to choose the radius of the disc used to estimate the curvature.
Briefly said, too large disc does not detect localised changes in curvature in the sense that it can capture more than one interface, on the other hand, too small disc has large error due to discretisation.
All realisation images in our simulation study are in resolution of $400 \times 400$ pixels (units).
Some recommendations on how to choose a suitable radius can be found in~\cite{bullard:1995}.
Based on these conditions and personal consultation with Mat\v{e}j L\'ebl (Institute of Information Theory and Automation, Czech Academy of Science), we use the radii $r=3$ and $r=5$.
Note that the results of the simulation study are very similar for both radii.
The presented histograms show the $p$-values of the tests using $r=5$ (with one exception mentioned below).

\begin{figure}
    \centering
    \includegraphics[width=15cm]{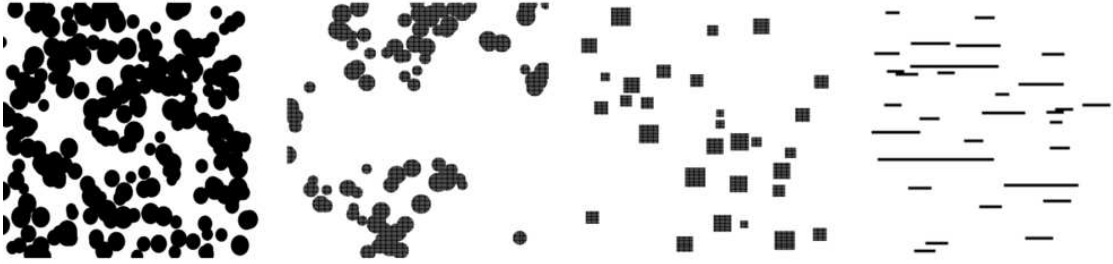}
    \caption{Example of realisation of the Boolean, the reduced Boolean, the square and the rectangle model, respectively.}
    \label{fig:BBrSNr}
\end{figure}

Examples of realisations of the illustrating models are shown in Fig.~\ref{fig:BBrSNr}.
The first picture is a realisation of the random disc Boolean model used in previous studies in~\cite{debayle:2021}, \cite{gotovac:2019}, \cite{gotovac:2021} and \cite{gotovac:2016} (see the last paper for details about parameters of the model).
The second realisation is simulated so that in (another) realisation of the Boolean model, each connected component is deleted with probability 1/2. 
It is called the reduced Boolean model in the sequel.
The third realisation is formed by disjoint squares whose ratio of the perimeter and the area comes from the same distribution as the ratio for the Boolean model (namely from the empirical distribution obtained from 100 realisations of Boolean model). 
We call it the square model in the sequel.
The fourth realisation is simulated as the process of disjoint rectangles with the same distribution of perimeters as the square perimeters, while one side has fixed length of 4 pixels (note that in this case, we use the disc with radius $r=3$ only for estimating the boundary curvature).
It is called the rectangle model in the sequel.

\begin{figure}
    \centering
    \includegraphics[width=15cm]{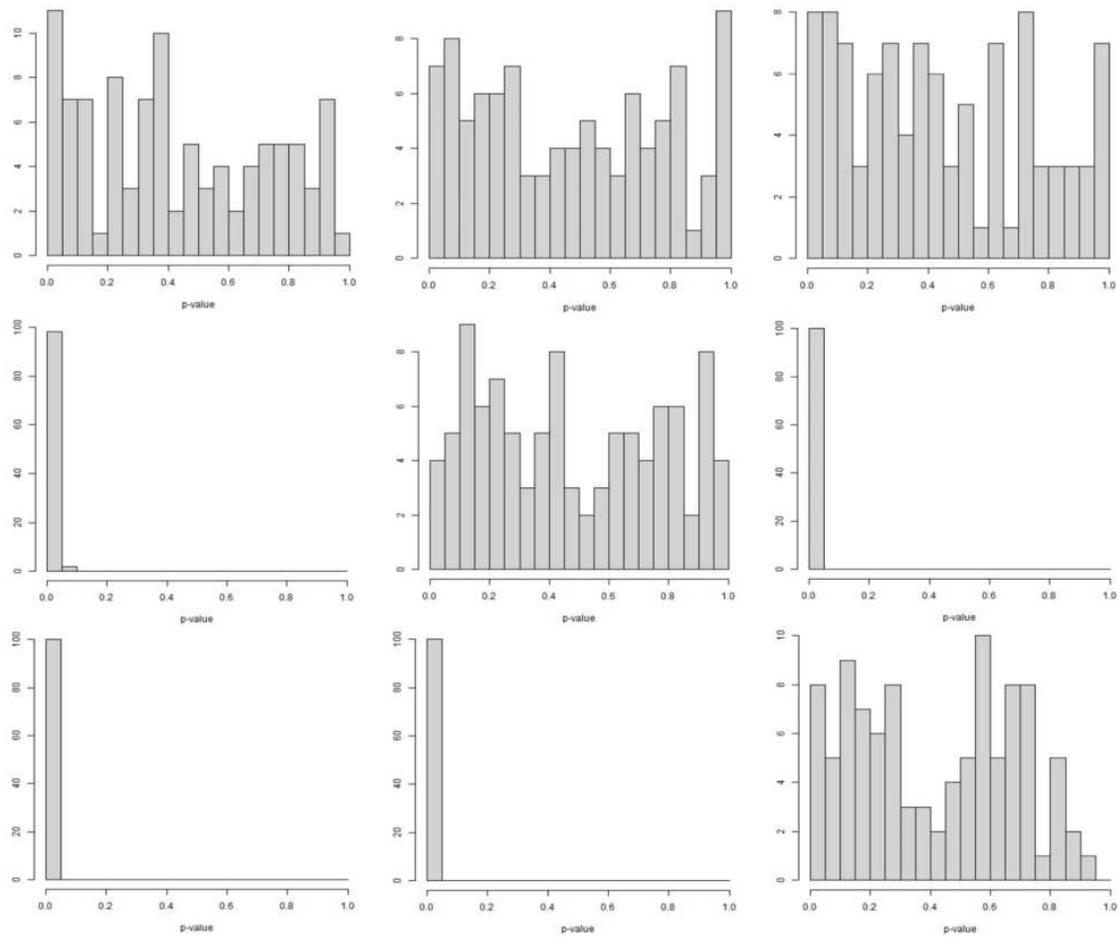}
    \caption{Histograms of $p$-values when testing the Boolean model vs the reduced Boolean model (the first row), the Boolean model vs the square model (the second row) and the square morel vs the rectangle model using both characteristics, i.e. the curvatures of the boundary and the ratios of the area and perimeter of the components (the first colomn), only the ratios (the second columns) and only the curvatures (the third columns).}
    \label{fig:BBrSNrhist}
\end{figure}

We want to show that the method does not distinguish between the Boolean model and reduced Boolean model since it is based only on the similarity of the components, but it distinguishes between the Boolean models and the square model due to the boundary curvature, as well as between the square model and the rectangle model due to the ratios of the perimeter and the area of the components. 
Indeed, we can see it in the first column of Fig.~\ref{fig:BBrSNrhist}. 
The histogram of $p$-values shows approximately uniform distribution when comparing the Boolean model and reduced Boolean model, but the $p$-values are very close to zero when testing the Boolean model vs the square model and the square model vs the rectangle model.
We moreover test the equality in distribution of the curvature functions and of the ratios of perimeters and areas separately.
The histograms of the $p$-values of the test for the ratios are shown in the second column, and the $p$-values of the test for the curvature functions are shown in the third column.
It is natural that the the $p$-values are approximately uniformly distributed for both tests of the Boolean model vs reduced Boolean model.
The meaning of the procedure is clear from other histograms. 
There, we can observe the equal distributions of the ratios of perimeters and areas, but clearly different distributions of the curvature functions when comparing the Boolean model and the square model, and conversely the agreement of the distribution of the curvature functions and different distributions of the ratios of perimeters and areas when testing the square model vs the rectangle model.
Thus, when we want to distinguish between realisations with differently shaped components, both characteristics must be taken into account.

\begin{figure}
    \centering
    \includegraphics[width=15cm]{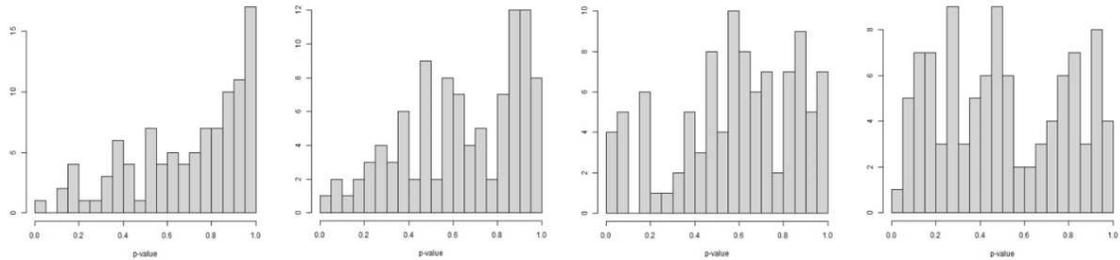}
    \caption{Histograms of $p$-values when testing the Boolean model vs the Boolean model using, respectively, 50, 30, 20 and 10 components from each realisation.}
    \label{fig:dependence}
\end{figure}

Next thing we observe in our simulation study is that when we have all components of the realisation in the sample, the $p$-values are greater than we expect.
It is seen when comparing the same models. 
The histograms of $p$-values are located to the right, while they become more uniform when we sample less components for testing, see Fig.~\ref{fig:dependence}. 
It is the effect of dependence of components in each realisation.
In realisations with densely placed components, the shape of one component affects the shape of another one, so they form something like a puzzle.
Such sets of components are then more similar then sets formed by independent components.
From histograms in Fig.~\ref{fig:dependence}, we conclude that in our case, we can take a sample of 10 components from realisation of Boolean model to eliminate the effect of dependence of the components.

\begin{figure}
    \centering
    \includegraphics[width=15cm]{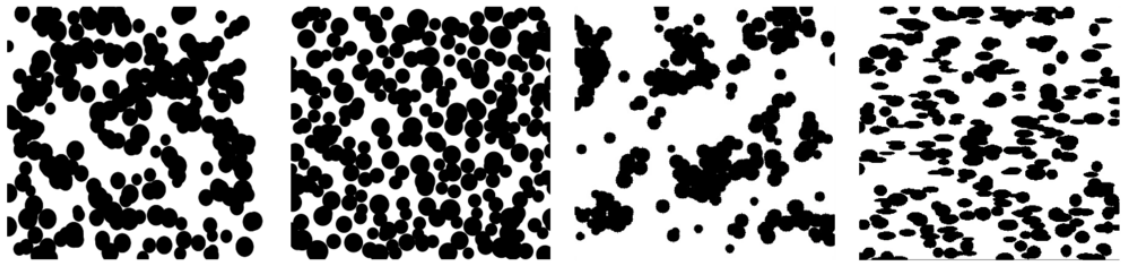}
    \caption{Example of realisation of the Boolean, the repulsive, the cluster and the ellipse model, respectively.}
    \label{fig:BCR}
\end{figure}

Further, we consider models compared before in~\cite{debayle:2021}, \cite{gotovac:2019}, \cite{gotovac:2021} and \cite{gotovac:2016}.
Except the Boolean model mentioned above, which appears in all the mentioned publications, we consider a model of partially repulsive particles (called the repulsive model in the sequel) and a model of particles forming clusters (called the cluster model in the sequel). 
Realisations of both these models are simulated as realisations of the random disc Quermass-interaction process~\cite{moeller:2008} with suitable chosen parameters.
More details about the parameters can be found in~\cite{gotovac:2016} and \cite{gotovac:2019}.
Note that the same repulsive model is employed for simulation studies in all above mentioned papers, similarly as the Boolean model, while the same cluster model is used only in~\cite{gotovac:2019}.
In~\cite{debayle:2021},  \cite{gotovac:2021} and \cite{gotovac:2016}, another cluster model is considered, which is not suitable for our current study since its realisations consist of too few components.
The fourth model is the Boolean model with grains to be ellipses (called the ellipse model in the sequel).
It appears, similarly as the cluster model, only in~\cite{gotovac:2019}, because in the other papers, its application would not be interesting.
Examples of realisations of the four models are shown in Fig.~\ref{fig:BCR}.

\begin{figure}
    \centering
    \includegraphics[width=15cm]{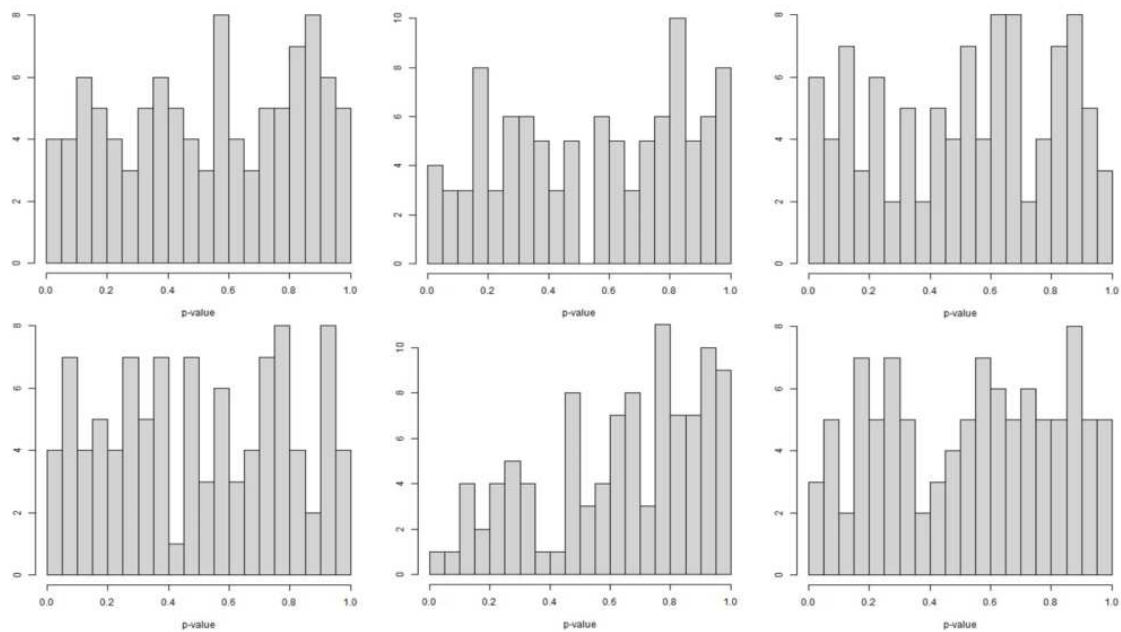}
    \caption{Example of realisation of the Boolean, the reduced Boolean, the square and the rectangle model, respectively.}
    \label{fig:the_same_hist}
\end{figure}

First, we test the similarity of the same models. 
For the Boolean model, we can see in Fig.~\ref{fig:dependence} that the $p$-values are approximately uniformly distributed for samples of the size between 10 and 20 components. 
Such a large sample can be viewed as a sample of weakly dependent components.
Therefore, we make samples of 10 and 20 components from each realisation of the remaining models.
The histograms in Fig.~\ref{fig:the_same_hist} shows that the $p$-values are uniformly distributed when testing the samples of 20 components for the repulsive model and for the ellipse model, while for the cluster model, the sample is not rare enough, it has uniformly distributed $p$-values for the samples of 10 components.

\begin{figure}
    \centering
    \includegraphics[width=15cm]{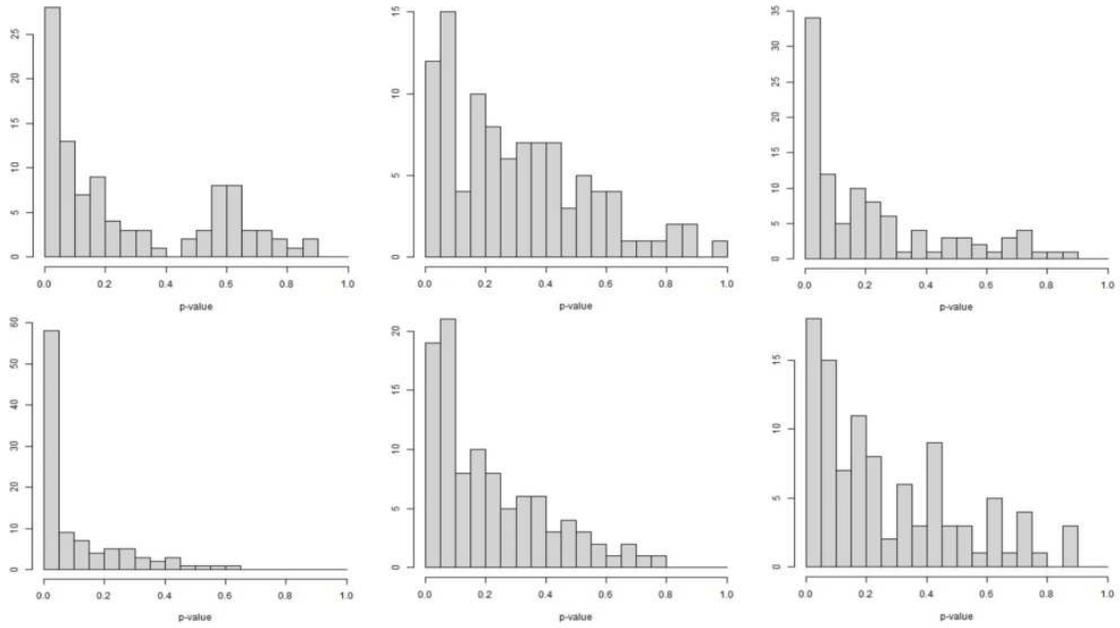}
    \caption{Histograms of $p$-values when testing similarity of the Boolean model vs the repulsive model (upper left), the Boolean model vs the cluster model (upper middle), the repulsive model vs the cluster model (upper right), the ellipse model vs the Boolean model (lower left), the ellipse model vs repulsive model (lower left) and the ellipse model vs the cluster model (lower left) using the samples of 10 components.}
    \label{fig:different_hist}
\end{figure}

Based on this observation, we use the samples of 10 components for testing similarity of different models. 
Their histograms are shown in~Fig.~\ref{fig:different_hist}.
We can see that the $p$-values are more or less close to zero, but the rejection of the similarity hypothesis is not very convincing.

\begin{figure}
    \centering
    \includegraphics[width=15cm]{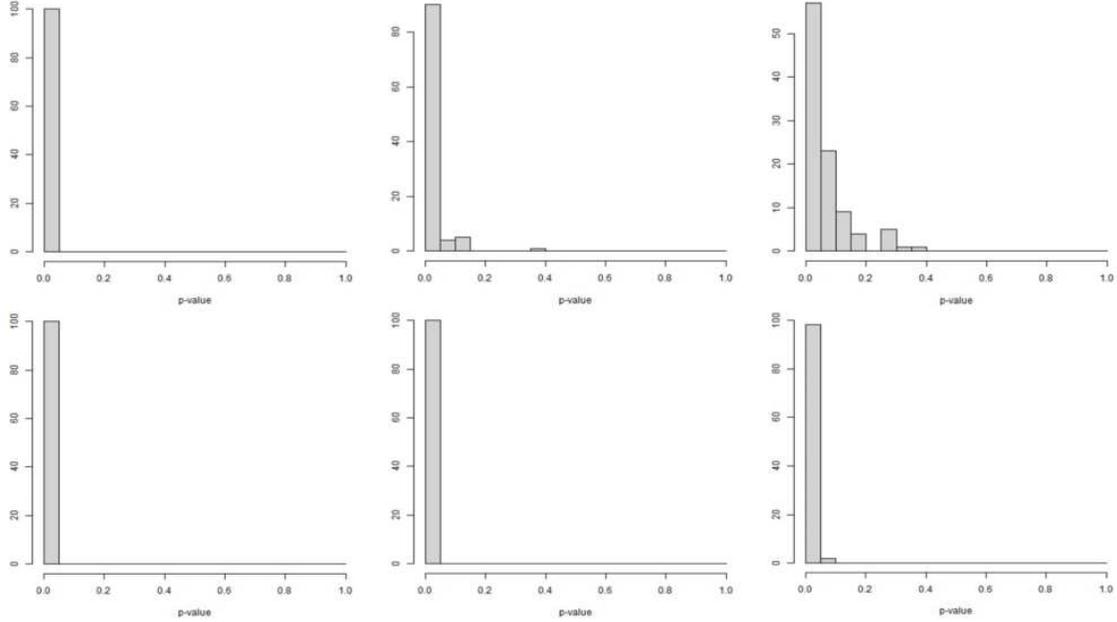}
    \caption{Histograms of $p$-values when testing similarity of the Boolean model vs the repulsive model (upper left), the Boolean model vs the cluster model (upper middle), the repulsive model vs the cluster model (upper right), the ellipse model vs the Boolean model (lower left), the ellipse model vs repulsive model (lower left) and the ellipse model vs the cluster model (lower left) using boot strap method and the samples of 100 components.}
    \label{fig:different_hist_100}
\end{figure}

We assume that this is due to the small number of components in the test sample, but we cannot create a larger sample from our simulated realisations because of the interdependence of the components. 
Therefore, we try to apply the boot strap method.
We mix all the components from each model together and, when testing the similarity of the two models, we randomly select 100 components of one model and 100 components of the second one and calculate the $p$-value of the similarity test.
We repeat this approach one hundred times to get 100 $p$-values for construction of histogram.
The histograms are shown in Fig.~\ref{fig:different_hist_100}.
We can see that except comparing the repulsive model and the cluster model, almost all $p$-values are less than 0.05 now.
The reason for larger $p$-values in the case of the repulsive model and the cluster model is the fact that many components in these models are formed by isolated discs that come from the same distribution, see the model parameters in~\cite{gotovac:2016}.

\section{Application to real data}

\begin{figure}
\centering
    \includegraphics[width=15cm]{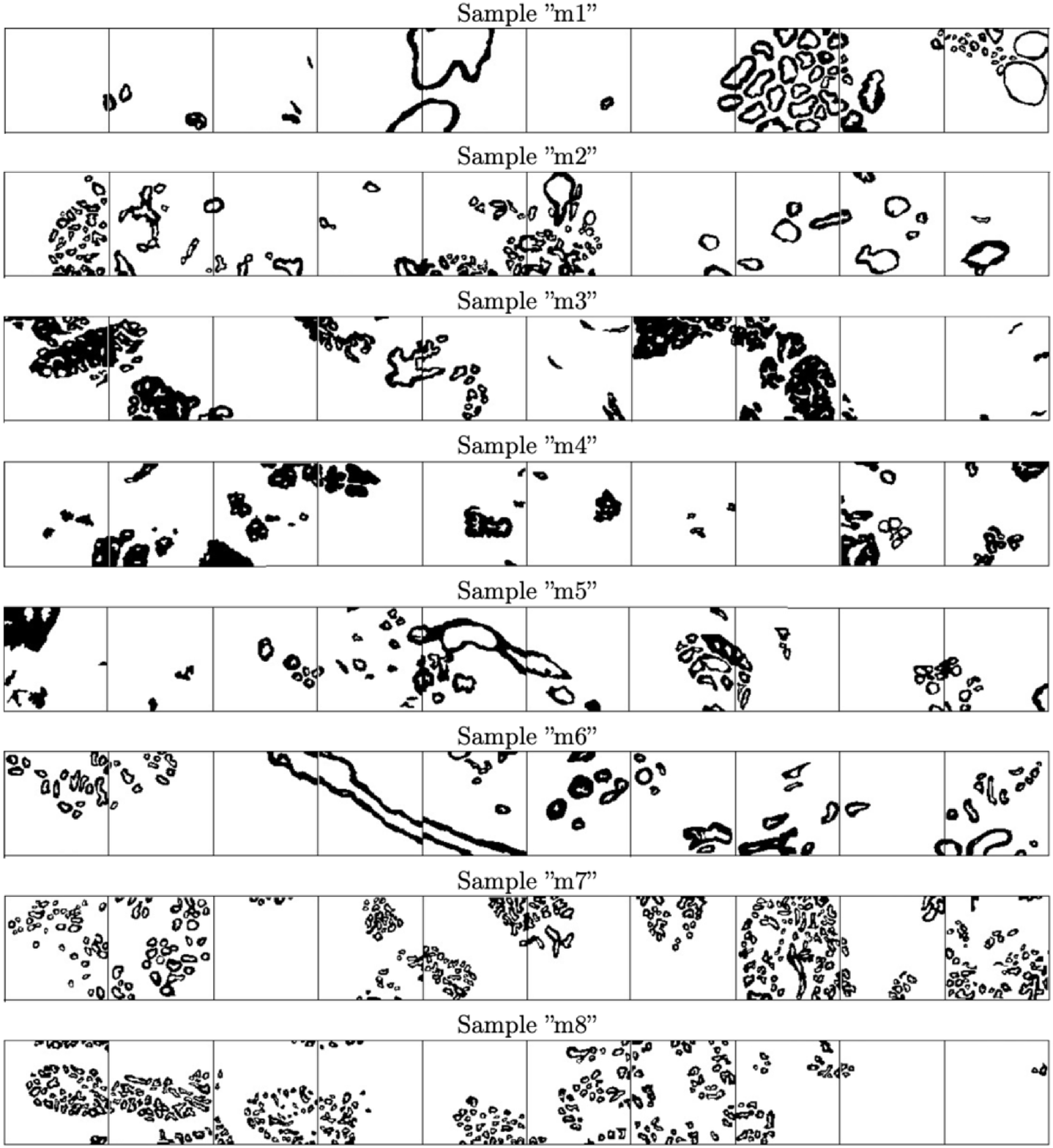}
\caption{Samples of mastopathy breast tissue.}
\label{fig:masto}
\end{figure}

\begin{figure}
 \centering
    \includegraphics[width=15cm]{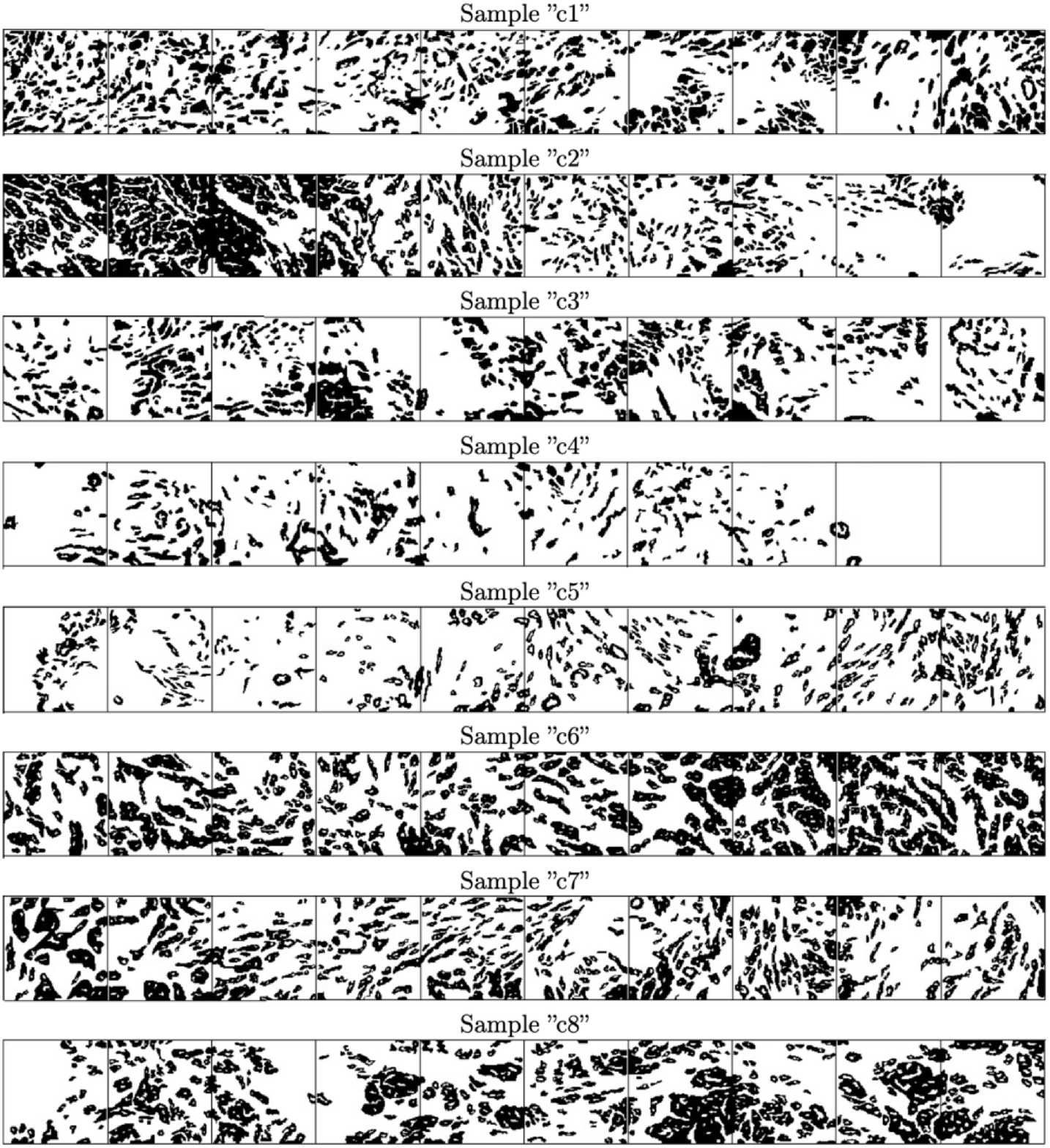}
\caption{Samples of mammary cancer.}
\label{fig:mamca}
\end{figure}

Finally, we apply the procedure to real data kindly provided by the authors of~\cite{mrkvicka:2011}.
We work with binary images of two different types of mammary tissue, namely 8 images of mastopathy tissue and 8 images of mammary cancer, see Fig.~\ref{fig:masto} and \ref{fig:mamca}.
Note that the breast contains a branching system of ducts spanning down from the nipple to glands.  
The tissue between the ducts and glands is made of fat and fibrous tissue of different proportions. 
The morphology of this tissue may indicate various malignant or benign changes.

The images are in resolution of 512 $\times$ 5120 pixels.
From each image, we randomly sample 20 components, and then, we test the similarity of the random sets represented by the images each to each, using $R=5$.
We repeat the procedure 100 times, while we evaluate the mean $p$-value of the test for each couple of images (including the image with itself) and the number of $p$-values below 0.05 which indicate significant dissimilarity on the classical level.
The results are introduced in Tab.~\ref{tab:meanp} and Tab.~\ref{tab:lowp}.
We can observe that the $p$-values are significantly lower and the number of $p$-values below 0.05 is significantly higher when comparing pairs of different types of tissue than that ones for pairs of the same types of tissue.
Just note that we repeated the same approach also for $R=3$ and for the samples of 10 components from each image and the results were very similar.

\begin{table}
\caption{Mean $p$-values (rounded to 2 decimal places) when comparing the corresponding samples 100 times. The values related to couples of different types of tissue are marked with italic font.}
\label{tab:meanp}
\begin{tabular}{cllllllllllllllll}
\textbf{} &   \textbf{m1} &   \textbf{m2} &   \textbf{m3} &   \textbf{m4} &   \textbf{m5} &   \textbf{m6} &   \textbf{m7} &   \textbf{m8} &   \textbf{c1} &   \textbf{c2} &   \textbf{c3} &   \textbf{c4} &   \textbf{c5} &   \textbf{c6} &   \textbf{c7} &   \textbf{c8} \\
\textbf{m1} & .81 & .46 & .35 & .02 & .19 & .41 & .04 & .04 & {\it .01} & {\it .02} & {\it .04} & {\it .10} & {\it .02} & {\it .03} & {\it .04} & {\it .06} \\
\textbf{m2} &     & .60 & .10& .00    & .05 & .38 & .07 & .03 & {\it .01} & {\it .04} & {\it .06} & {\it .20} & {\it .03} & {\it .05} & {\it .06} & {\it .11} \\
\textbf{m3} &     &     & .87 & .35 & .57 & .20& .00 & .00 & {\it .00} & {\it .00} & {\it .01} & {\it .01} & {\it .00} & {\it .01} & {\it .00} & {\it .00} \\
\textbf{m4} &     &     &     & .90 & .13 & .04 & .00 & .00 & {\it .00} & {\it .00} & {\it .00} & {\it .00} & {\it .00} & {\it .01} & {\it .00} & {\it .00} \\
\textbf{m5} &     &     &     &     & .78 & .32 & .00 & .00 & {\it .00} & {\it .00} & {\it .01} & {\it .01} & {\it .00} & {\it .01} & {\it .00} & {\it .01} \\
\textbf{m6} &     &     &     &     &     & .65 & .03 & .01 & {\it .01} & {\it .03} & {\it .03} & {\it .14} & {\it .01} & {\it .09} & {\it .04} & {\it .10} \\
\textbf{m7} &     &     &     &     &     &     & .59 & .49 & {\it .14} & {\it .16} & {\it .10} & {\it .36} & {\it .45} & {\it .02} & {\it .19} & {\it .15} \\
\textbf{m8} &     &     &     &     &     &     &     & .61 & {\it .10} & {\it .19} & {\it .10} & {\it .28} & {\it .40} & {\it .01} & {\it .15} & {\it .13} \\
\textbf{c1} &     &     &     &     &     &     &     &     & .53 & .38 & .32 & .17 & .27 & .11 & .35 & .26 \\
\textbf{c2} &     &     &     &     &     &     &     &     &     & .55 & .43 & .41 & .35 & .17 & .50 & .37 \\
\textbf{c3} &     &     &     &     &     &     &     &     &     &     & .57 & .29 & .18 & .31 & .47 & .38 \\
\textbf{c4} &     &     &     &     &     &     &     &     &     &     &     & .57 & .25 & .20& .35 & .40\\
\textbf{c5} &     &     &     &     &     &     &     &     &     &     &     &     & .55 & .03 & .35 & .16 \\
\textbf{c6} &     &     &     &     &     &     &     &     &     &     &     &     &     & .54 & .24 & .40\\
\textbf{c7} &     &     &     &     &     &     &     &     &     &     &     &     &     &     & .51 & .42 \\
\textbf{c8} &     &     &     &     &     &     &     &     &     &     &     &     &     &     &     & .54
\end{tabular}
\end{table}

\begin{table}
\caption{The number of $p$-values bellow $.05$ when comparing the corresponding samples 100 times. The values related to couples of different types of tissue are marked with italic font.}
\label{tab:lowp}
\begin{tabular}{cllllllllllllllll}
\textbf{} &   \textbf{m1} &   \textbf{m2} &   \textbf{m3} &   \textbf{m4} &   \textbf{m5} &   \textbf{m6} &   \textbf{m7} &   \textbf{m8} &   \textbf{c1} &   \textbf{c2} &   \textbf{c3} &   \textbf{c4} &   \textbf{c5} &   \textbf{c6} &   \textbf{c7} &   \textbf{c8} \\
\textbf{m1} &     0   & 4 & 9 & 92 & 29 & 5 & 81 & 81 & {\it 97} & {\it 92} & {\it 82} & {\it 57} & {\it 88} & {\it 79} & {\it 80} & {\it 70} \\
\textbf{m2} &     & 3 & 67 & 100 & 71 & 13 & 71 & 84 & {\it 95} & {\it 86} & {\it 78} & {\it 37} & {\it 86} & {\it 65} & {\it 74} & {\it 57} \\
\textbf{m3} &     &     &   0  & 5 & 1 & 29 & 99 & 100 & {\it 100} & {\it 100} & {\it 98} & {\it 98} & {\it 100} & {\it 96} & {\it 99} & {\it 98} \\
\textbf{m4} &     &     &     &  0   & 47 & 77 & 100 & 100 & {\it 100} & {\it 100} & {\it 100} & {\it 100} & {\it 100} & {\it 96} & {\it 100} & {\it 100} \\
\textbf{m5} &     &     &     &     &  0   & 19 & 100 & 100 & {\it 100} & {\it 100} & {\it 97} & {\it 96} & {\it 100} & {\it 95} & {\it 99} & {\it 93} \\
\textbf{m6} &     &     &     &     &     &   0  & 87 & 96 & {\it 96} & {\it 91} & {\it 89} & {\it 54} & {\it 97} & {\it 62} & {\it 84} & {\it 62} \\
\textbf{m7} &     &     &     &     &     &     & 1 & 5 & {\it 40} & {\it 36} & {\it 64} & {\it 16} & {\it 6} & {\it 91} & {\it 29} & {\it 50} \\
\textbf{m8} &     &     &     &     &     &     &     & 1 & {\it 52} & {\it 29} & {\it 60} & {\it 15} & {\it 17} &{\it  97} & {\it 33} & {\it 50} \\
\textbf{c1} &     &     &     &     &     &     &     &     & 6 & 12 & 19 & 43 & 20 & 63 & 13 & 28 \\
\textbf{c2} &     &     &     &     &     &     &     &     &     & 2 & 13 & 8 & 20 & 52 & 3 & 14 \\
\textbf{c3} &     &     &     &     &     &     &     &     &     &     & 2 & 18 & 47 & 18 & 5 & 12 \\
\textbf{c4} &     &     &     &     &     &     &     &     &     &     &     & 1 & 21 & 39 & 20 & 12 \\
\textbf{c5} &     &     &     &     &     &     &     &     &     &     &     &     & 1 & 89 & 23 & 44 \\
\textbf{c6} &     &     &     &     &     &     &     &     &     &     &     &     &     & 7 & 35 & 16 \\
\textbf{c7} &     &     &     &     &     &     &     &     &     &     &     &     &     &     & 2 & 12 \\
\textbf{c8} &     &     &     &     &     &     &     &     &     &     &     &     &     &     &     & 3
\end{tabular}
\end{table}

\section{Conclusion}

A new statistical test for assessing (dis)similarity of two random sets has been constructed.
It works with two realisations - one realisation of each of the random sets.
The procedure focuses only on shapes of the components of the random sets, namely on the curvature of their boundaries together with the ratios of their perimeters and areas, and it does not take into account the positions of the components in the realisations, since it is very often required in practical applications.

The described method is equipped by a simulation study.
The study shows that under quite mild conditions, the test has large power when distinguishing realisations of different models.
The power is larger than the method in~\cite{gotovac:2016} and one of the methods in~\cite{gotovac:2019}, which also distinguishes realisations based on the shape of the components, and comparable to another method in~\cite{gotovac:2019} and to the method in~\cite{debayle:2021}, which, however, take into account the placement of components.
Another advantage with respect to the method in~\cite{gotovac:2016} is that it is not heuristic, and moreover, it does not require a lot of input parameters as needed in~\cite{gotovac:2016} and~\cite{debayle:2021}.

A small complication occurs when components of compared realisations are too close to each other, because the method assumes independence of the components, which is not satisfied when the components are too densely placed.
However, the problem has an easy solution - only a sample of the components can be considered instead of all components.
In this case, the dependence is reduced and the method works very well.

Finally, the procedure is applied to real data.
The data consists of pictures of two different types of tissue, namely the tissue of mammary cancer and mastopathic tissue.
We have 8 pictures of each type.
The aim is to distinguish between different types and to assess the pictures of the same type of the tissue as similar.
Considering how different the images of the same type appear to be at first glance and how difficult it is to identify specific distinguishing features for different types, the results of the method are very good.

From the above observations, we can conclude that the new method works very well and has a high potential to be a useful tool for comparing realisations of random sets.

\subsubsection*{Acknowledgment}

Supported by The Czech Science Foundation, project No.~19-04412S.


\begin{thebibliography}{00}

\bibitem{bullard:1995} Bullard J.~W., Garboczi E.~J., Carter W.~C., and Fuller E.~R. Jr.: 
Numerical methods for computing interfacial mean curvature,
Computational Materials Science, vol. 4, pp. 103–116, 1995.

\bibitem{chiu:2013} S.~N.~Chiu, D.~Stoyan, W.~S.~Kendall, and J.~Mecke, Stochastic geometry and its applications. John Wiley \& Sons, New York, 2013.




\bibitem{debayle:2021} J. Debayle, V. Gotovac Doga\v{s}, K.~Helisov\'a, J.~Stan\v{e}k, and M.~Zikmundov\'a ``Assessing similarity of random sets via skeletons,'' Methodology and Computing in Applied Probability, DOI: 10.1007/s11009-020-09785-y,


\bibitem{gotovac:2019}  V.~Gotovac, ``Similarity between random sets consisting of  many components.'' Image Anal Stereol, vol.38, pp.~185--99, 2019.

\bibitem{gotovac:2021} V.~Gotovac Doga\v{s} and K.~Helisov\'a, ``Testing equality of distributions of random convex compact sets via theory of~N-distances,'' Methodology and Computing in Applied Probability, 2021+, DOI: 10.1007/s11009-019-09747-z


\bibitem{gotovac:2016}  V.~Gotovac, K.~Helisov\'a, and I.~Ugrina, ``Assessing dissimilarity of random sets through convex compact approximations, support functions and envelope tests,'' Image Anal Stereol, vol.~35, pp.~181--93, 2016.

\bibitem{gretton:2012} Gretton A, Borgwart KM, Rash MJ, Scholkopf B, Smola A (2012) A Kernel Two-Sample Test. J Mach Learn Res 13: 723--73



\bibitem{hermann:2015} P.~Hermann, T.~Mrkvi\v{c}ka, T.~Mattfeldt,  M.~Min\'{a}rov\'{a}, K.~Helisov\'{a}, O.~Nicolis, F.~Wartner, and M.~Stehl\'{i}k,
``Fractal and stochastic geometry inference for breast cancer: a case
study with random fractal models and Quermass-interaction process,'' Stat in Med, vol. 34.18, pp.~2636--61, 2015.

\bibitem{klebanov:2006} L.~B.~Klebanov, $\mathcal{N}$-distances and their applications.  Karolinum Press, Charles University, Prague, 2006.







\bibitem{matheron:1975} G.~Matheron, Random Sets and Integral Geometry. John Wiley \& Sons Inc, New-York, 1975.

\bibitem{molchanov:2005} I.~Molchanov, Theory of random sets. Springer, New York, 2005.

\bibitem{moeller:2008} J.~M\o{}ller and K.~Helisov\'{a}, ``Power diagrams and Interaction processes for unions of discs,'' Adv Appl Probab, vol.~40, pp.~321--47, 2008.

\bibitem{moeller:2010} J.~M\o{}ller and K.~Helisov\'{a}, ``Likelihood inference for unions of interacting discs,'' Scand Stat, vol.~37, pp.~365--81, 2010.

\bibitem{mrkvicka:2011} Mrkvi\v{c}ka T, Mattfeldt T (2011): Testing histological images of mammary tissues on compatibility with the Boolean model of random sets. Image Analysis and Stereology 30(1), 11--18.

\bibitem{myllymaki:2017} M.~Myllym\"aki, T.~Mrkvi\v{c}ka, P.~Grabarnik, H.~Henri Seijo, and U.~Hahn, ``Global envelope tests for spatial processes,'' Journal of the Royal Stat Society: Series B (Statistical Methodology), vol.~79, pp.~381--404, 2017.

\bibitem{neumann:2016} M.~Neumann, J.~Stan\v{e}k,  O.~M.~Pecho, L.~Holzer, V.~Bene\v{s}, and V.~Schmidt, ``Stochastic 3D modeling of complex three-phase microstructures in SOFC-electrodes with completely connected phases,'' Comp Mat Sci vol.~118, pp.~353--64, 2016.


\bibitem{serra:1982}
J.~Serra, Image Analysis and Mathematical Morphology, vol.2: Theoretical Advances. Academic Press, 1982.


\end{thebibliography}
\end{document}